\documentclass{epl2}
\usepackage{amsmath}
\usepackage{amssymb}
\usepackage{bm}
\usepackage{color}
\usepackage{epsfig}
\usepackage{graphicx}

\title{Pulse-pumped double quantum dot with spin-orbit coupling.}
\shorttitle{Pulse-pumped double quantum dot} 
\author{D.V. Khomitsky\inst{1} \and E.Ya. Sherman \inst{2,3}}
\shortauthor{D.V. Khomitsky, E.Ya. Sherman}
\institute{\inst{1} Department of Physics, University of Nizhny Novgorod, 23
Gagarin Avenue, 603950 Nizhny Novgorod, Russian Federation \\
\inst{2} Department of Physical Chemistry, Universidad del Pa\'is
Vasco, 48080 Bilbao, Spain \\
\inst{3} IKERBASQUE Basque Foundation for Science, Alameda Urquijo 36-5,
48011, Bilbao, Bizkaia, Spain}
\pacs{73.63.Kv}{Quantum dots (electronic transport)}
\pacs{72.25.Dc}{Spin polarized transport in semiconductors}
\pacs{72.25.Pn}{Current-driven spin pumping}
\abstract{We consider the full driven quantum dynamics of a qubit 
realized as spin of electron in a one-dimensional 
double quantum dot with spin-orbit coupling.
The driving perturbation is taken in the form of a single half-period 
pulse of electric field. Spin-orbit coupling leads to a nontrivial 
evolution in the spin and charge densities making the dynamics 
in both quantities irregular. As a result, the charge density distribution 
becomes strongly spin-dependent. The transition from the field-induced tunneling 
to the strong coupling regime is clearly seen in the charge and spin channels.   
These results can be important for the understanding of the 
techniques for the spin manipulation in nanostructures.}
\begin{document}
\maketitle
Short title: Pulse-pumped double quantum dot
\pagebreak

\textit{Introduction.} Driven qubits (quantum two-level systems) 
attract attention due to the richness of phenomena occurring
in different regimes of coupling to external field \cite{Kohler05}
and possible applications for quantum information devices. 
Spins of electrons in semiconductor quantum dots (QDs) are widely expected as a possible realization of 
qubits.\cite{Burkard99} In the presence of spin-orbit (SO) coupling even simple
systems such as the single-electron QDs show a rich dynamics in 
coupled charge and spin channels. The spin dynamics in relatively weak electric
fields in a single QD is well understood in terms of the 
electric dipole spin resonance (EDSR). In this process the electric 
field at the frequency matching the Zeeman resonance for electron spin in magnetic field drives
the orbital motion, and as a result, causes a spin-flip. 
This coupling is in the core of the proposal by Rashba and Efros \cite{Rashba1,Rashba2} 
for a new technique to manipulate the spin states
by an external electric field. The EDSR spin-flip rate
is orders of magnitude greater than the
rate of the transitions due to the magnetic component
of electromagnetic field. The efficiency of EDSR for
GaAs QDs has been proven experimentally in Ref.\cite{Nowack} where
the electric-field induced spin Rabi oscillations have been
observed, demonstrating the abilities of a coherent spin manipulation. 
Another approach was employed in Ref.\cite{Pioro}, where the
electric field was used for driving electrons in a nonuniform magnetic
field, thus causing a spin dynamics. The coherent spin responce
shows that the problem with the spin dephasing in GaAs
QDs, also arising due to the SO coupling,\cite{Semenov,Fabian05} can be overcome.

A generic system to study the driven behavior is a single parabolic
one-dimensional QD where the electron states are the states
of the harmonic oscillator. The coordinate and
momentum can be expressed with the ladder operators satisfying
known commutation relations. As a result, well-known selection
rules for coordinate and momentum matrix elements can be employed 
and the consideration can be sufficiently simplified.
However, completely isolated QDs cannot be used for quantum
information applications since an interdot interaction is necessary to
produce and manipulate many-body states and a large scale of the
system is required to make it work. Moreover, the experiments
necessarily use at least a double QD \cite{Nowack,Pioro}
to detect the driven spin state relative to the spin of the reference electron,
and the role of the induced interdot tunneling on the spin Rabi oscillations has been
noticed.\cite{Nowack} This ``photon-induced'' tunneling can put a 
limitation on the abilities to manipulate the spins.

The information on the coupled quantum dynamics of orbital and spin degrees 
of freedom when both are nontrivial, is, however, scarce. 
Here we make a step forward and address full driven spin/charge quantum dynamics in a 
one-dimensional double QD \cite{Sanchez06,Zulicke08,Ulloa06,Wu} 
by considering a pumped motion of electron. These systems lavished
attention due to their deceptive simplicity, where
despite a well-established Hamiltonian, the rich variety of phenomena 
can be observed.
In the classical regime, where only over-the-barrier motion can occur,
the dynamics was studied in Ref.\cite{Khomitsky} and revealed interesting
coupled irregular behavior of the spin and charge motion. In the more 
experimentally interesting quantum regime, the 
tunneling between single QDs plays the crucial
role in the low-energy states, while the higher-energy
ones are extended at a larger spatial scale. 
In the presence of SO coupling the
electron states and the interdot tunneling become spin-dependent. 
For the driven systems this spin-dependence will be seen below.
Nonlinear systems driven by a time-dependent field can show a strongly
irregular chaos-like behavior.\cite{Lin92,Sipe,driven} However, the existence of the quantum chaos,
in contrast to the classical one, is disputed due to a finite set of
energy eigenstates of the system. At the same time,
the interplay of irregularities for spin and charge degrees of 
freedom is important for the entire system dynamics: the driven motion being not 
chaotic, is, however, strongly irregular. 
These irregularities, both in the 
orbital and the spin motion, limiting the
abilities of the spin manipulations, will be also of our interest in this paper.

\begin{figure}[tbp]
\centering
\includegraphics[scale=0.7]{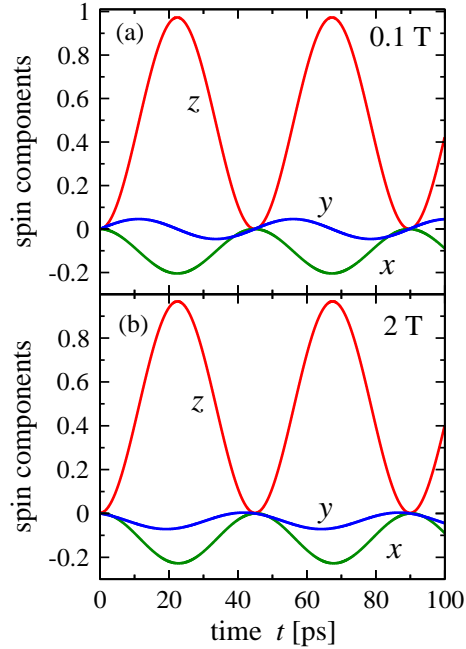}
\caption{Free evolution of $\sigma_{R}^{i}(t)$ for
all spin components, as marked near the plots for the initial spin-up case: 
(a) for weak field $H_z=0.1$ T (here we use $\alpha =1.0\cdot 10^{-9}$ eVcm, $\beta=0$), (b)-relatively strong field $H_z=2$ T.
The probability to find the electron in the right dot $w_{R}(t)$ almost coincides
with $\sigma_{R}^{z}(t)$, and only a slight spin deviation 
from the initial value due to SO coupling is observed.}
\label{free}
\end{figure}

\textit{Hamiltonian and time evolution.} We describe a one-dimensional
double QD by the Hamiltonian, similar to suggested in Ref.\cite{Romano}, for the particle with mass $m$
in a quartic potential 
\begin{equation}
U(x)=U_{0}(-2(x/d)^{2}+(x/d)^{4}),
\end{equation}
where the minima located at $d$ and $-d$ are separated by a barrier of
height $U_{0}$. We assume that the barrier is opaque enough to
ensure that the ground state is close to the even linear combination of the
oscillator states located near the minima. The frequency of
the single-dot harmonic oscillator is determined by 
$\omega _{0}=2\sqrt{2U_{0}}/d\sqrt{m}$ and the corresponding oscillator length $a_{0}=%
\sqrt{\hbar /m\omega _{0}}$. The interdot semiclassical tunneling
probability in the ground state: 
\begin{equation}
w_{t}=\exp \left[ -\frac{8\sqrt{2}}{3}\frac{\sqrt{mU_{0}}d}{\hbar }\right]
\end{equation}
is small. The tunneling splits the ground state with the energy close to 
$-U_{0}+\hbar\omega_{0}/2$ by $\Delta E_{g}$ of the order of 
$\hbar\omega_{0}w_{t}.$ The high-energy states with the orbital quantum number $%
n\gg 1$ can be treated semiclassically for the potential 
$U(x)=U_{0}(x/d)^{4},$ yielding the eigenenergy: 
\begin{equation}
E_{n}=\left(
\frac{3\Gamma^2(3/4)}{2\sqrt{2\pi}}
\right)^{4/3}
n^{4/3}\hbar\omega_{0}\left(\frac{\hbar\omega_{0}}{U_{0}}\right)^{1/3},
\end{equation}
where $\Gamma$ is the gamma function. Correspondingly, the maximum semiclassical momentum achieved when the electron
passes the vicinity of the $x=0$ point, is $p_{n}^{\max }\sim \sqrt{2mE_{n}}%
\sim n^{2/3}\left(\hbar\omega/U_{0}\right)^{1/6}\hbar/a_{0}.$

\begin{figure}[tbp]
\centering
\includegraphics[scale=0.7]{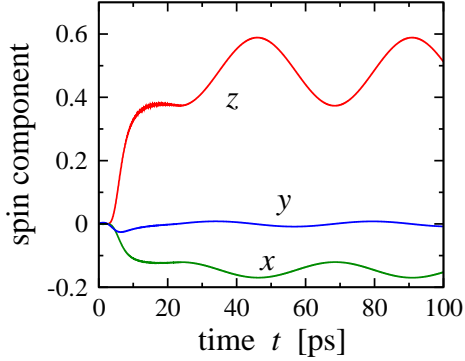}
\caption{Evolution of the spin components (as marked near the plots) for initial
spin-up state in a weak magnetic field $H=0.1$ T, where the spin is influenced only by
the SO coupling. 
As in fig.\ref{free}(a), here we use $\alpha =1.0\cdot 10^{-9}$ eVcm, $\beta=0$. 
The external field is weak, $f=1/8$,  and the pulse duration $T=25$ ps.}
\label{ev:H01}
\end{figure}

For the system in a static magnetic field $H_z$ and driven by an external 
electric field $\mathcal{E}(t)$ directed along the $x$%
-axis, the Hamiltonian is the sum of three terms $H=H_{0}+H_{\mathrm{so}}+%
\widetilde{V}$ with: 
\begin{eqnarray}
H_{0} &=&\frac{p_{x}^{2}}{2m}+U(x)+\frac{g}{2}\mu _{B}\sigma _{z}H_{z}
\label{ham} \\
H_{\mathrm{so}} &=&\left( \beta \sigma _{x}+\alpha \sigma _{y}\right)
p_{x},\qquad \widetilde{V}=-e\mathcal{E}(t)x,
\end{eqnarray}
where $e$ is the electron charge, and $\sigma_{i}$ are the Pauli matrices. 
The effect of  $H_{z}$
is described by the Zeeman term only, where $g$ is the Land\'e factor. The
SO interaction is the sum of bulk-originated Dresselhaus $\left(\beta \right)$ 
and structure-related Rashba $\left( \alpha \right) $ terms.
The corresponding velocity operator 
\begin{equation}
v\equiv \dot{x}=\frac{{\rm i}}{\hbar }[H_{0}+H_{\mathrm{so}},x]={p_{x}}/{m}+\beta
\sigma _{x}+\alpha \sigma _{y}.
\label{velocity}
\end{equation}
is spin-dependent.

To describe the driven motion we use the following approach. 
First, we diagonalize exactly the Hamiltonian $H_{0}+H_{\rm so}$ in the truncated basis
of spinors $\psi _{n}(x)\left|\sigma\right>$ with corresponding
eigenvalues $E_{n\sigma}$ and obtain the new basis set 
$\left|\psi_{\mathbf n}\right>$ where bold $\mathbf{n}$ 
incorporates the corresponding spin index. The first four low-energy
states are then defined as: $\left|\psi_{\mathbf 1}\right>=\psi_{1}(x)\left|\uparrow\right>$,
$\left|\psi_{\mathbf 2}\right>=\psi_{1}(x)\left|\downarrow\right>$,
$\left|\psi_{\mathbf 3}\right>=\psi_{2}(x)\left|\uparrow\right>$, and
$\left|\psi_{\mathbf 4}\right>=\psi_{2}(x)\left|\downarrow\right>$. 

Since the energy $E_{ns}$ increases 
as $n^{4/3}$, the truncated basis is sufficient for the purposes we consider
below. Then, we build in this basis the matrix of the Hamiltonian $\widetilde{V}$ 
and study the full dynamics with the wavefunctions in the form: 
\begin{equation}
\left| \Psi \right\rangle =\sum_{\mathbf{n}}\xi _{\mathbf{n}}(t)e^{-{\rm i}E_{%
\mathbf{n}}t/\hbar }\left| \psi _{\mathbf{n}}\right\rangle.
\end{equation}
The time dependence of $\xi_{\mathbf{n}}(t)$ with $\widetilde{V}=-e\mathcal{E}(t)x$
is calculated as: 
\begin{equation}
\frac{d}{dt}\xi _{\mathbf{n}}(t)={\rm i}\frac{e}{\hbar }\mathcal{E}(t)\sum \xi _{%
\mathbf{m}}(t)x_{\mathbf{nm}}e^{-{\rm i}\left( E_{\mathbf{m}}-E_{\mathbf{n}%
}\right) t/\hbar },
\label{maineq}
\end{equation}
where $x_{\mathbf{nm}}\equiv \left\langle\psi_{\mathbf{n}}\right|\widehat{x}\left|\psi_{\mathbf{m}}\right\rangle$. 
Since we do not assume a periodic driving field, we do not
resort to the Floquet method \cite{Wu,Shirley}, but use the direct numerical calculation
instead. Taking into account that we are interested in a relatively short-term dynamics, we 
neglect momentum relaxation. This approximation is allowed at low temperatures 
and low excitation energies. 

The spin-dependence of the matrix element of coordinate responsible for the 
spin dynamics can be seen from expression ${\rm i}({E_{\mathbf{n}}-E_{\mathbf{m}}})x_{\mathbf{nm}}$
$={\hbar }\left\langle\psi_{\mathbf{n}}\right|\widehat{v}\left|\psi_{\mathbf{m}}\right\rangle,$
and the spin-dependent terms in the velocity in (\ref{velocity}).

\textit{Observables and results.} Our main goal is to calculate the probability- 
\begin{equation}
\rho (x,t)=\Psi^{\dagger}(x,t)\Psi (x,t)\label{charge}
\end{equation}
and spin- 
\begin{equation}
S_{i}(x,t)=\Psi^{\dagger}(x,t)\sigma_{i}\Psi(x,t),  \label{Sz}
\end{equation}
density using (\ref{maineq}). With these distributions 
we find the gross quantities for the right QD: 
\begin{eqnarray}
w_{R}(t) &=&\int_{0}^{\infty }\rho (x,t)dx, \\
\sigma _{R}^{i}(t) &=&\int_{0}^{\infty }S_{i}(x,t)dx,
\end{eqnarray}
where $w_{R}(t)$ is the probability to find electron and $%
\sigma _{R}^{i}(t)$ is the expectation value of spin component. 
As the electron wavefunction at $t=0$ we
take linear combinations of four low-energy states. 
The initial state is localized in the left QD, and we choose
combinations: $\xi_{\mathbf 1}=\xi_{\mathbf 3}=1/\sqrt{2}$ for the spin-up,
and  $\xi_{\mathbf 2}=\xi_{\mathbf 4}=1/\sqrt{2}$ for the spin-down states. 

We consider a structure with $d=25\sqrt{2}$ nm and $U_{0}=10$ meV, with the corresponding
oscillator length $a_{0}=23$ nm.  The four lowest spin-degenerate energy levels are
$E_{1}=3.938$ meV, $E_{2}=4.030$ meV, $E_{3}=9.782$ meV, $E_{4}$=11.590 meV counted from the bottom of a single QD. 
The tunneling splitting $\Delta E_{g}=E_{2}-E_{1}=0.092$ meV. 
The parameters of the SO coupling are assumed to be
$\alpha =1.0\cdot 10^{-9}$ eVcm, $\beta =0.3\cdot 10^{-9}$ eVcm.
We assume the magnetic field $H_{z}=2$ T, which produces the 
Zeeman spin splitting of the levels 
$\Delta_{z}=E_{n\downarrow }-E_{n\uparrow}=0.054$ meV for the
Land\'e-factor of electron in GaAs $g=-0.45$.

In numerical calculations the basis of 20 states with the energies up to 42 meV was 
employed. We consider the period of time evolution $0\le t\le 100 $ ps tracking the system
in the transient and following stationary processes.

\begin{figure}[tbp]
\centering
\includegraphics[scale=0.7]{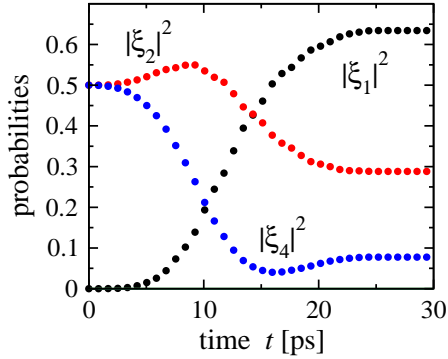}
\caption{Evolution of the components of the wave function (as marked near the plots) for initial
spin-down state.  The external field is characterized by $f=1/2$, and the pulse duration $T=25$ ps.}
\label{ev:xi}
\end{figure}

We begin with the analysis of the free evolution, where $\widetilde{V}=0$.
The results are presented in fig.\ref{free} for two magnetic fields. 
One can see the free tunneling Rabi oscillations with the period $T_R$
of approximately 50 ps and the corresponding spin oscillations. 
Figure \ref{free} shows that no spin flip can be achieved in the tunneling at this system parameters.
Moreover, $w_{R}(t)$ (not shown in the figure) and $\sigma _{R}^{z}(t)$  almost coincide, confirming that
tunneling by itself does not lead to a considerable spin dynamics. The similarity of the
free motion pictures for $H_z=0.1$ and $H_z=2$ T shows that the $z-$axis field, even producing
the Zeeman splitting comparable with the tunneling splitting of the ground state, does not
modify the tunneling. In the tunneling, spin precession mainly around the $y$-axis due to
the Rashba SO term is observed. The precession angle defined for this
case as $\phi={\rm atan}\left(\sigma_{R}^{y}/\sigma_{R}^{z}\right)$ achieves the maximum
$\phi_{\max}\approx 0.25$ at $t=T_{R}/2$. This number should be compared 
with the precession angle for the classical motion, where displacement of electron
by the distance $l$ leads to the precession angle $\phi_{\rm cl}(l)=2ml\alpha/\hbar^2$. 
For given geometry of the structure with $l=2d$, one obtains $\phi_{\rm cl}(2d)=1.2$, 
considerably larger than $\phi_{\max}$. This implies that combination of tunneling
and SO coupling cannot be treated semiclassically being strongly different
from what can be expected. Namely, due to the enhanced role of the tunneling the 
precession angle becomes considerably smaller.
Here a general remark is appropriate. It was suggested \cite{Baz} and analyzed \cite{Buttiker,Sokolovski} how
to use the Larmor spin precession in magnetic field as the measure of the time electron 
spends inside a barrier. In the presence of SO coupling every
electron has an associated momentum-dependent magnetic field acting on the spin.
Whether the spin precession in this field can be used as a measure of tunneling time 
is an interesting question. 

\begin{figure}[tbp]
\centering
\includegraphics[scale=0.7]{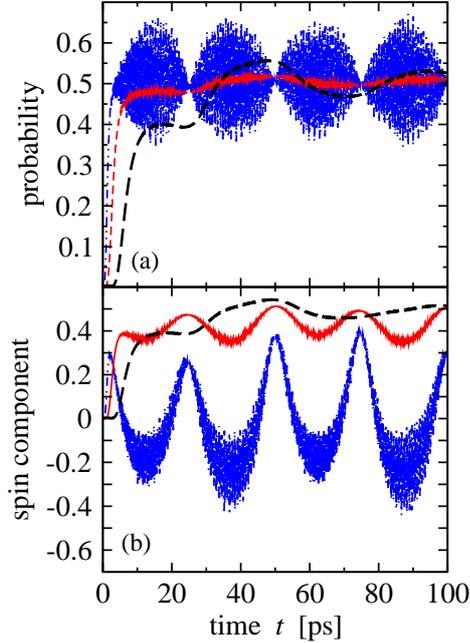}
\caption{Evolution of the probabilities $w_R(t)$ (a) and $\sigma_{R}^{i}(t)$ (b) under
the electric field pulse with $f=1/8$, $f=1/2$, and $f=2$ for the initial spin-up state.
The pulse durations $T=25$ ps. Dashed line shows the $f=1/8$ regime. With the increase in the pulse field, the motion becomes 
less regular, and more high-energy component contribute into the dynamics. }
\label{ev:spinup}
\end{figure}

As the external perturbation we consider half-period electric field pulses in the form $%
\mathcal{E}(t)=\mathcal{E}_{0}\sin (\pi t/T)$ for $0<t<T$. For the pulse duration we
assume $T=T_{R}/2.$ The spectral width of the pulse covers both the 
spin and the tunneling splitting of the  ground states, thus, driving the
spin and orbital dynamics simultaneously. Chosen pulse duration satisfies the 
condition $T\omega_{0}\gg 1$, that
is the corresponding frequencies are much less than the energy difference
between the orbital levels corresponding to a single dot, and, therefore, the
high energy states follow the perturbation adiabatically. The field
strength is characterized by parameter $f$ such that $e\mathcal{E}_{0}\equiv f\times U_{0}/2d$.
For the coupling we consider three regimes: (i) a relatively weak field ($f\ll 1$) 
where the shape of the quartic potential 
remains almost intact, and the tunneling is still crucially important,
(ii) intermediate filed,  and (iii) strong field $f>1$, 
where the perturbation already considerably changes the
low-energy part the spectum. 

As the first example, we consider the pulse-driven evolution in a weak
magnetic field $H_z=0.1$ T, where the spin dynamics is determined mainly by the
SO coupling. The dynamics is shown in fig.\ref{ev:H01}. Even a weak field strongly changes
the observables: in the rich transient dynamics 
it quickly almost equilazes the probabilities to find electron 
in the right and left dot, and the subsequent 
oscillations of probability around 0.5 value occur. Same behavior is demonstrated
by the spin components - they weakly oscillate around mean value since the
electron travels a shorter distance during the density oscillations, however,
a similar to fig.\ref{free} spin polarization  $\sigma_{R}^{i}(t)/w_{R}(t)$
is achieved. 

\begin{figure}[tbp]
\centering
\includegraphics[scale=0.7]{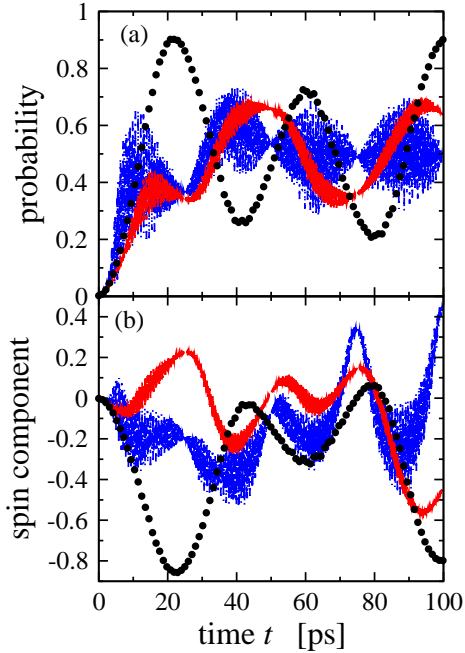}
\caption{Evolution of the probabilities $w_R(t)$ (a) and $\sigma_{R}^{i}(t)$ (b) under
the electric field pulse with $f=1/8$, $f=1/2$, and $f=2$ for the initial spin-down state.
Circle symbol line corresponds to the $f=1/8$ regime. 
The pulse durations $T=25$ ps. Similarly to fig.\ref{ev:spinup}, with the increase in the pulse field, 
the motion becomes less regular, and more high-energy component contribute into the dynamics.
However, the dynamics is much richer here. For the $f=1/2$ case the main 
underlying $|\xi_{\mathbf n}|^2$ are shown in fig.\ref{ev:xi}.}
\label{ev:spindown}
\end{figure}

We present driven dynamics in the Hilbert
space with the coefficients $\xi_{\mathbf n}$ for $f=1/2$ in fig.\ref{ev:xi}.
This picture shows only the largest remaining after the end
of the pulse components, demonstrating the orbital and spin dynamics,
equilibrating the probably density as $|\xi_{\mathbf 4}|^2$ decreases
with respect to $|\xi_{\mathbf 2}|^2$, and causing the spin evolution 
by increasing  $|\xi_{\mathbf 1}|^2$. 

The observables for the pulses with different ${\mathcal E}_{0}$
are shown in fig.\ref{ev:spinup} and fig.\ref{ev:spindown}
for the initial spin-up and spin-down states. 
Comparison of fig.\ref{ev:spinup} and fig.\ref{ev:spindown} shows that
the time-dependence of the spin density, and, more important, of
the probability is quite different for different initial spin states despite
the fact that in both cases the dynamics consists of low-
frequency motion corresponding to the spectrum of the four
low-energy states superimposed by the set
of very-high frequency oscillations due to the  
involvement of the higher-energy states. For the
spin-up initial state, the weak and moderate external field just equalizes the
probability for the electron to occupy the left and the right
dots, and then only high-frequency oscillations occur. 
For the initial spin-down state, the dynamics is much richer:
the low-frequency oscillations with a large amplitude are 
clearly seen at long times. In both cases the amplitude of the
fast oscillations increases with the increase in the field. 
The difference between these two regimes is 
non-trivial and can be understood as follows. The coupled spin-charge 
dynamics is considerably determined by the difference in the energy of the nearest spin up and spin
down states with opposite parity. The larger is the distance, the less efficient
is the SO coupling for the coupled dynamics  both in the low-
and high-frequency domains. For the low-energy spin-up states 
it is $\left| E_{\mathbf 4}-E_{\mathbf 1}\right|=\Delta E_{g}+\Delta_{z}$, 
while for the initial spin-down state it is  
$\left|E_{\mathbf 3}-E_{\mathbf 2}\right| =\Delta E_{g}-\Delta _{z}$. 
For $H_{z}=2$ T, $\left(\Delta E_{g}-\Delta_{z}\right)/\left(\Delta E_{g}+\Delta_{z}\right)\approx 0.25$, 
making the spin flip at the given
spectrum of the pulse, for the latter case easier, and the spin dynamics richer. 

\begin{figure}[tbp]
\centering
\includegraphics[scale=0.35]{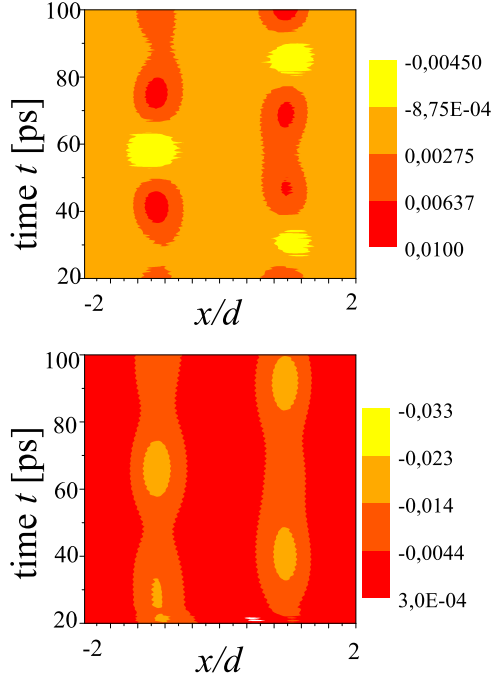}
\caption{Evolution of the  spin densities under the strong electric
field $f=2$ pulse  for two pulse durations $T=12$ ps (upper panel) and $T=25$ ps
(lower panel). The variation in the pulse duration leads to the different outcome for
the dominating final $S_z$ of the transferred electron.  }
\label{ev:2D}
\end{figure}

To have a better insight into the evolution of the spin density, we
illustrate the dynamics by a plot of $S_{z}(x,t)$ in fig.\ref{ev:2D} 
for  pulse durations $T=12$ and $T=25$ ps, $f=2$ and the initial spin-down
state. The figure shows that for the shorter $T=12$ ps pulse, the $S_z$ component in the 
right dot $(x=d)$ is positive at most of the time after the end of the pulse, 
illustrating the controlled spin flip accompanying the electron transfer.
By moving along the time axis at $x=d$ we follow oscillations of the
$S_z$ component consistent with the free tunneling on the scale of $T_R$.
As for the evolution under the longer pulse $T=25$ ps shown in the lower panel
of fig.\ref{ev:2D}, the shape of $S_{z}(x,t)$ remains qualitatively the same,
however the details of the distribution in the right dot are different. Namely, the
dominating parts of the profile correspond to the negative $S_z$ at
all times at $x\approx d$. This remarkable difference 
between the results for two pulse durations 
provides a good illustration for the interplay of
a driven spin-and-charge motion and the tunneling occurring simultaneously.
Namely, the stronger and the longer the electric pulse is, the more overall changes it provides
for the coupled charge and spin dynamics. Since the spin evolution is the rotation,
a too short (or a too weak) electric pulse may "under-rotate" the spin with
respect to the desirable state, as we have seen earlier for the weak pulses. 
A too strong (or a too long) electric pulse can "over-rotate" the
spin, which may pass the desired flipped state in the neighboring dot. This observation
illustrated in fig.\ref{ev:2D} suggests to carefully control and adjust both the
strength and duration of the electric pulses required for the controlled spin manipulation
in heterostructures with significant SO coupling.

\textit{Conclusions.} We have studied the full driven quantum spin and charge
dynamics of a spin qubit in one-dimensional double single-electron QD
with SO coupling. Equations of motion in a pulsed electric field have been solved numerically 
exactly in a finite basis set in the regimes of relatively weak, moderate, 
and relatively strong coupling. The dynamics is strongly determined by the
pulse amplitude and duration. The results show that the time dependence of the 
spin and charge distributions are strongly irregular with the spin-dependent 
evolution in the charge density pattern. Electron spin flip  can be achieved 
at certain pulse durations and amplitudes. 
These conclusions emphasize the importance of the SO coupling which should be
taken into account in experimental and technological realizations of spin-based 
qubits. 

D.V.K. is supported by the RNP Program of Ministry of 
Education and Science RF (Grants No. 2.1.1.2686, 2.1.1.3778, 2.2.2.2/4297, 2.1.1/2833), 
by the RFBR (Grant No. 09-02-1241-a), by the USCRDF (Grant No. BP4M01), 
by "Researchers and Teachers of Russia" FZP Program NK-589P,
and by the President of RF Grant No. MK-1652.2009.2. EYS is 
supported by the University of Basque Country UPV/EHU grant
GIU07/40 and MCI of Spain grant FIS2009-12773-C02-01.   
The authors are grateful to L.V. Gulyaev for technical assistance.

\end{document}